\documentclass[
superscriptaddress,
aps,prxquantum,
reprint,
twocolumn,
amsmath,amssymb,showpacs
]{revtex4-2}

\usepackage{amsmath,mathtools}
\usepackage{mathrsfs}
\usepackage{ulem}
\usepackage{changes}
\usepackage{amsfonts}
\usepackage{physics}
\usepackage{txfonts}
\usepackage{dsfont}
\usepackage{amssymb}
\usepackage{graphicx}
\usepackage{float}
\usepackage{xcolor} 
\usepackage{changes}
\usepackage{url}
\usepackage[colorlinks=true, citecolor=blue, urlcolor=blue]{hyperref}
\usepackage{array}

\begin{document}

\title{Universal learning of nonlocal entropy via local correlations in non-equilibrium quantum states}

\author{Hao Liao}
\affiliation{National Engineering Laboratory on Big Data System Computing Technology, Guangdong Province Engineering Center of
China-made High Performance Data Computing System, College of Computer Science and Software Engineering, Shenzhen
University, Shenzhen 518060, China}

\author{Xuanqin Huang}
\affiliation{National Engineering Laboratory on Big Data System Computing Technology, Guangdong Province Engineering Center of
China-made High Performance Data Computing System, College of Computer Science and Software Engineering, Shenzhen
University, Shenzhen 518060, China}

\author{Ping Wang}
\email{wpking@bnu.edu.cn}
\affiliation{Faculty of Arts and Sciences, Beijing Normal University, Zhuhai 519087, China}

\date{\today}

\begin{abstract}
Characterizing the nonlocal nature of quantum states is a central
challenge in the practical application of large-scale quantum computation
and simulation. Quantum mutual information (QMI), a fundamental nonlocal
measure, plays a key role in quantifying entanglement and has become
increasingly important in studying nonequilibrium quantum many-body
phenomena, such as many-body localization and thermalization. However,
experimental measurement of QMI remains extremely difficult, particularly
for nonequilibrium states, which are more complex than ground states.
In this Letter, we employ a multilayer perceptron (MLP) to establish
a universal mapping between the QMI and local correlations only up
to second order for nonequilibrium states generated by quenches in
a one-dimensional disordered XXZ model. Our approach provides a practical
method for experimentally extracting QMI, readily applicable in platforms
such as superconducting qubits. Moreover, this work will establishes
a general framework for reconstructing other nonlocal observables,
including Fisher information and out-of-time-ordered correlators.
\end{abstract}

\maketitle

%A key difference of quantum mechanism from classical mechanism is the non local feature of the quantum state, full access to which requires exponential number local measurements. A representative non-local quantity is the quantum Rényi entropy(QRE), which has been widely applied in the entanglement detection, characterization of phase transition and open quantum system. In recent years, the behavior of non-equilibrium dynamics of the quantum Rényi entropy is also predicted to be relevant to non-equilibrium many body phase induced by disorder and hence the precise measurement of these non-local quantity not only has pivotal application to quantum information theory but also be useful to understand the fundamental quantum non-equilibirum statistics, quantum chaotic and quantum gravity.

\textit{Introduction.}\textbf{---}A fundamental distinction between
quantum and classical mechanics lies in the non-local character of
quantum states~\cite{BrunnerRMP2014}. Their complete characterization
generally requires an exponential number of local measurements, presenting
a major challenge for quantum technologies~\cite{MacFarlanePTORSL2003}.
A paradigmatic non-local measure is the quantum Rényi entropy (QRE),
widely used in entanglement detection~\cite{GuehnePR2009,IslamNature2015},
the characterization of phase transitions~\cite{HeylROPIP2018,LambertPRA2005},
and studies of many-body physics~\cite{EisertNP2015}. Recent work
has further revealed that non-equilibrium dynamics of the QRE~\cite{NanduriPRB2014}
exhibits signatures of disorder-induced many-body localization (MBL)~\cite{SinghNJP2016}.
Accurate measurement of such non-local quantities is therefore crucial
not only to quantum information theory, but also to understanding
fundamental aspects of quantum non-equilibrium statistics~\cite{KaufmanScience2016,Venderleyprl2018,LukinScience2019},
quantum chaos, and quantum gravity~\cite{RovelliPRL1996}.

However, reconstructing non-local QRE in large-scale quantum systems
remains challenging, as it generally requires quantum state tomography
(QST), which becomes infeasible due to the exponential growth of Hilbert
space dimension and measurement overhead. Although QRE has been measured
in specific systems where multiple copies of a many-body state can
be prepared and interfered~\cite{DaleyPRL2012,AbaninPRL2012,KaufmanScience2016,LukinScience2019},
its efficient measurement in generic, large-scale systems remains
an open problem. Several approaches have recently been proposed to
overcome this fundamental scaling issue, including randomized measurements~\cite{BrydgesScience2019,ElbenPRL2020},
matrix product state tomography~\cite{LanyonNP2017}, neural network
quantum states~\cite{Carleo2017,TorlaiNP2018,CarrasquillaNMI2019},
quantum neural networks~\cite{ShinQIP2024}, many-body interference~\cite{KaufmanScience2016},
machine learning techniques~\cite{RiegerPRA2024,wuNC2024}, and classical
shadows~\cite{HuangNP2020}. These methods aim to predict non-local
observables using data that scale polynomially with system size~\cite{wuNC2024,RiegerPRA2024}.
Nevertheless, efficient and general reconstruction of QRE continues
to pose significant challenges due to various method-specific limitations.

In this Letter, we employ a multilayer perceptron (MLP) to learn a
universal function for predicting QMI in quench dynamics---a quantity
derived from quantum Rényi entropy---using only correlations with
orders less than two, without resorting to QST. In contrast to the
classical shadow method, randomized measurements{} and multi-task
networks, which rely on the complete set of measurement outcomes\cite{BrydgesScience2019,ElbenPRL2020,HuangNP2020}
or short-range data\cite{wuNC2024,RiegerPRA2024}, we instead construct
a mapping between the quantum mutual information and a substantially
more compact and experimentally accessible data set. This approach
is motivated by two key insights: (i) measurements of second-order
correlations are significantly more efficient than those of higher-order
correlations, which suffer from reduced signal-to-noise ratios and
increased experimental overhead, even though higher-order measurements
are feasible in certain systems~\cite{XuPRL2018}; (ii) although
QMI of a nondegenerate ground state can be determined from second-order
correlations alone~\cite{ChenPRA2012,QiQuantum2019,wuNC2024}, it
remains unclear whether non-equilibrium states in closed quantum many-body
systems permit such efficient reconstruction, given their far greater
complexity. Remarkably, we demonstrate efficient and accurate prediction
of disordered averaged QMI during long-time quench dynamics in the
one-dimensional XXZ model with random field---across both many-body
localized(MBL) and thermal phases---using only second-order correlations
and a single universal nonlinear function trained on one Hamiltonian
configuration, which can be readily adapted to predict the QMI across
diverse experimental quantum platforms.

\begin{figure}[ptb]
\includegraphics[width=1\columnwidth]{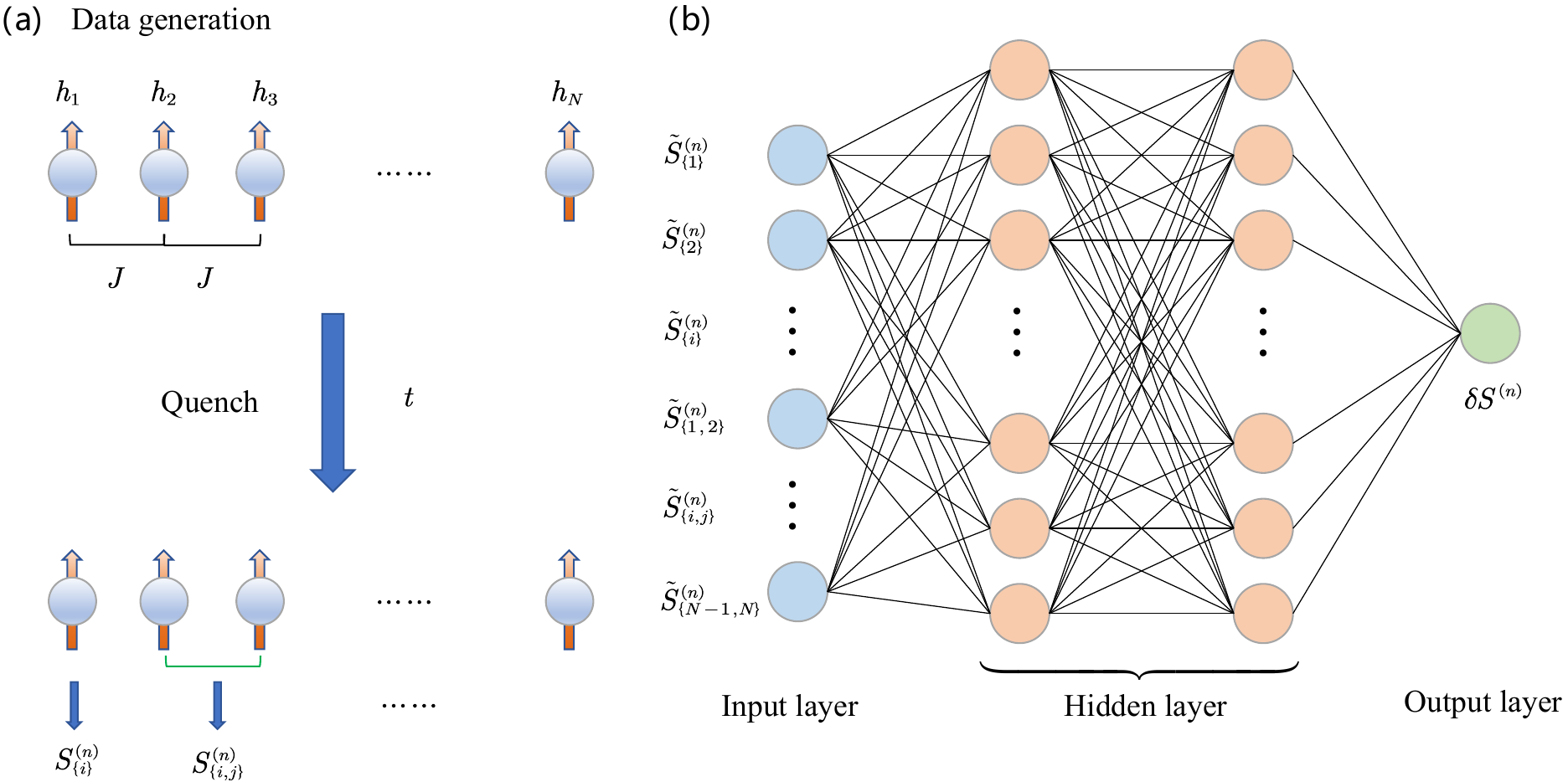} \caption{\textbf{The protocol to training the Rény entropy: }a. The quenching
dynamics of XXZ model and generation of local correlation and QMI
for non-equilibrium state; $\textbf{b}$. The MLP method to learn
the non-linear mapping between QMI $\delta S^{(n)}$ and local irreducible
entropy $\delta S_{i}^{(n)}$, $\delta S_{\{i,j\}}^{(n)}$ with $i,j$
be the index of the spins.}
\label{fig:protocol} 
\end{figure}

\textit{Model.---} We study the dynamics of QMI following a sudden
quench of duration $t$ in the disordered XXZ model---a paradigmatic
system for exploring many-body localization (MBL)~\cite{SierantROPP2025}.
The QMI, a highly non-local observable, is defined as 
\[
\delta S^{(n)}=S^{(n)}(\hat{\rho}_{A})+S^{(n)}(\hat{\rho}_{B})-S^{(n)}(\hat{\rho}),
\]
for suitably chosen subsystems $A$ and $B$. Here, $S^{(n)}(\hat{\rho})=\log_{2}\mathrm{Tr}\hat{\rho}^{n}/(n-1)$
is the Rényi entropy (with integer $n$), and $\hat{\rho}_{A/B}=\mathrm{Tr}_{B/A}[\hat{\rho}(t)]$
are reduced density matrices. The Hamiltonian of the XXZ chain is
given by 
\begin{equation}
\hat{H}=\sum_{i=1}^{N}h_{i}\hat{I}_{i,z}+\sum_{i=1}^{N-1}\left[\left(\hat{I}_{i,x}\hat{I}_{i+1,x}+\hat{I}_{i,y}\hat{I}_{i+1,y}\right)+J_{z}\hat{I}_{i,z}\hat{I}_{i+1,z}\right],\label{eq:Hamiltonian_MBL}
\end{equation}
where $\hat{I}_{i,\alpha}=\hat{\sigma}_{i,\alpha}/2$ (with $i=1,\dots,N$
and $\alpha=x,y,z$) is the spin operator at site $i$, and $\hat{\sigma}_{i,\alpha}$
are Pauli matrices. The disorder fields $h_{i}$ are independent random
variables uniformly distributed in $[-W,W]$, and $J_{z}$ controls
the anisotropy. For $J_{z}=1$, the system undergoes an MBL-to-thermal
transition at a critical disorder strength $W_{c}\approx3.6$~\cite{LuitzPRB2015,SerbynPRX2015}.

\begin{figure*}[ptb]
\includegraphics[width=1.8\columnwidth]{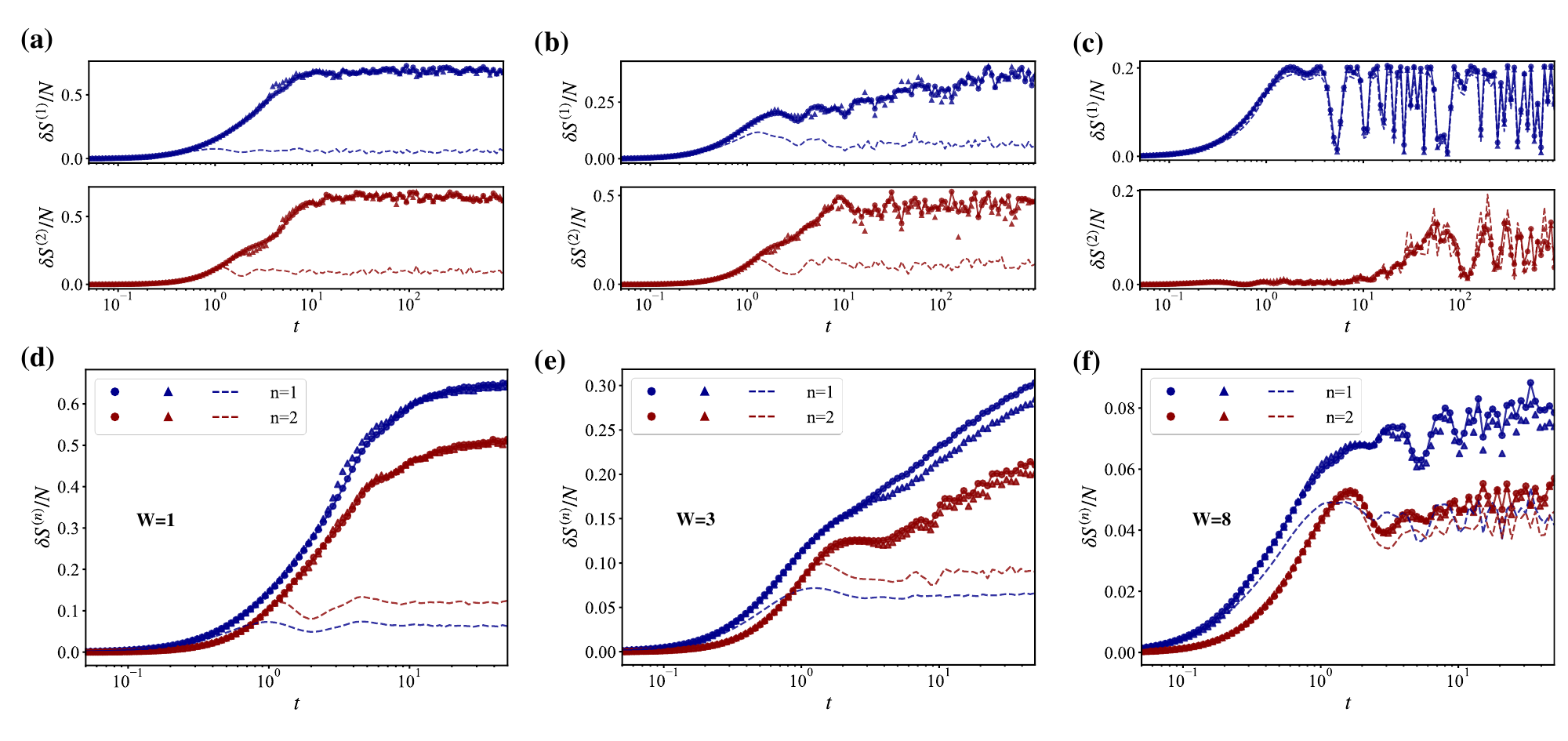} \caption{\textbf{Machine learning of the dynamics of QMI }(a), (b), (c): The
dynamics of QMI $\delta S^{(2)}/N$ of single disorder realization
for different disorder strength: (a). $W=1$, (b). $W=3$, (c). $W=8$
when fixing $J_{z}=1$ and $N=10$. $\delta S^{(2)}$ is calculated
for the partitions $A=\{1,2,3,4,5\}$ and $B=\{6,7,8,9,10\}$.\textbf{
}(d), (e), (f): The disorder averaged dynamics of $\delta S^{(n)}/N$
upper corresponding to the case (e), (d), (f). These results are obtained
by averaging over $100$ random disorder configuration. In all these
figures, the circles, triangle and dashed line denotes the result
that calculated by exact simulation, MLP and CCE method respectively.}
\label{fig:Comparision} 
\end{figure*}

As illustrated in Fig.~\ref{fig:protocol}(a), we initialize the
system in a random product state $|\Psi(0)\rangle=\bigotimes_{i=1}^{N}|m_{i}\rangle$,
where $m_{i}\in\{+,-\}$ and $|\pm\rangle_{i}$ denote the eigenstates
of $\hat{I}_{i,z}$. The state then evolves unitarily under the Hamiltonian~(\ref{eq:Hamiltonian_MBL})
for a duration $t$. For the resulting state $|\Psi(t)\rangle=e^{-i\hat{H}t}|\Psi(0)\rangle$,
we compute the QMI from the density matrix $\rho(t)=|\Psi(t)\rangle\langle\Psi(t)|$
for the chosen subsystems $A$ and $B$.

We then employ the MLP to learn the relationship between the non-local
QMI $\delta S^{(n)}$ and the local correlations (up to second order)
of the non-equilibrium quantum state $|\Psi(t)\rangle$ generated
by quenching dynamics described above. As is well known, the density
matrix of the quantum state $|\Psi(t)\rangle$ can be fully described
by the correlations $C_{\boldsymbol{\alpha}_{\boldsymbol{i}_{L}}}^{\boldsymbol{i}_{L}}=\langle\Psi\left(t\right)|\prod_{n=1}^{L}\hat{\sigma}_{\alpha_{i_{n}}}^{i_{n}}|\Psi\left(t\right)\rangle$
of all the clusters, where $L$ denotes the size of the cluster, $\boldsymbol{i}_{L}=\{i_{1},i_{2},\cdots,i_{L}\}$($1\le i_{n}\le N$)
is the index of spins, while $\boldsymbol{\alpha}_{\boldsymbol{i}_{L}}=\{\alpha_{i_{1}},\alpha_{i_{2}},\cdot\cdot\cdot,\alpha_{i_{L}}\}$($\alpha_{i_{n}}=x,y,z$)
is the collection of the spin indices. The total number of these correlations
is $4^{N}-1$. Measurement of these correlations is termed quantum
state tomography(QST) and full access of these correlations can in
principle give an exact reconstruction of the density matrix and hence
the QMI $\delta S^{(n)}$. However, it is in fact impossible to measure
all these correlations especially for quantum system with large size
$N$ due to exponentially increased measurement times. Consequently,
it is highly desirable to investigate the relation between the QMI
$\delta S^{(n)}$ and the local correlations up to the second order,
which are easy to measure experimentally.

A central question is whether a universal relation exists between
$\delta S^{(n)}$ and low-order correlations at arbitrary time $t$
in the XXZ model. If such a function is found, it would enable experimental
determination of the QMI $\delta S^{(n)}$ using only up to second-order
local correlations, drastically reducing the measurement overhead
required to reconstruct the QMI.

\textit{Method.---}Before discussing the machine learning method,
we first introduce the cluster correlation expansion method (CCE)
for reconstructing the QMI inspired by methods for calculating decoherence~\cite{YangPRB2008,YangPRB2009}
, as it provides physical insight into the underlying structure. Motivated
by the gradual buildup of correlations during quenching dynamics,
we can formally express the QMI as {[}see Supplemental Materials{]}
\begin{equation}
\delta S^{(n)}=-\sum_{L=2}^{N}\sum_{C_{L}}\tilde{S}_{C_{L}}^{(n)},\label{eq:CEM}
\end{equation}
where the sum runs over all connected clusters $C_{L}$ of size $L$
that intersect both subsystems $A$ and $B$ (i.e., $C_{L}\cap A\ne\emptyset$
and $C_{L}\cap B\ne\emptyset$). Clusters contained entirely within
$A$ or $B$ do not contribute. Here, $\tilde{S}_{C_{L}}^{(n)}$ denotes
the irreducible $n$-th Rényi entropy for cluster $C_{L}$, defined
recursively via\cite{YangPRB2008,YangPRB2009,RispoliNature2019} {[}see
supplementary materials{]} 
\begin{equation}
\tilde{S}_{C_{L}}^{(n)}=S_{C_{L}}^{(n)}-\sum_{M=1}^{L-1}\sum_{C'_{M}\subset C_{L}}\tilde{S}_{C'_{M}}^{(n)},\label{eq:renyi-entropy-2-1}
\end{equation}
with the base case $\tilde{S}_{C_{1}}^{(n)}=S_{C_{1}}^{(n)}$, and
where $S_{C_{L}}^{(n)}=\log_{2}\mathrm{Tr}[\rho_{C_{L}}^{n}]/(1-n)$
is the Rényi entropy of the reduced state on $C_{L}$. Each $S_{C_{L}}^{(n)}$
can be determined from correlations with order $\le L$. Although
Eq.~(\ref{eq:CEM}) is exact, its evaluation requires high-order
clusters, which becomes infeasible for large $L$. Alternately, we
can truncate $L$ to $L_{\mathrm{max}}$ and make the approximation
$\delta S^{(n)}\approx-\sum_{L=1}^{L_{\mathrm{max}}}\sum_{C_{L}}\tilde{S}_{C_{L}}^{(n)}$(called
CCE-$L_{\mathrm{max}}$) for short quenching time since the correlation
is gradually built order by order and hence the genuine QMI of high
order clusters should be very small for short quenching time. This
truncation only requires lower order correlations with order $\le L_{\mathrm{max}}$.
The simplified case $L_{\mathrm{max}}=2$ is the CCE-2 which only
needs the first order expectation and second order correlations. However,
these approximations obviously breaks down at longer times due to
the proliferation of higher-order correlations.

Motivated by the CCE method, we want to ask whether the QMI can still
be reconstructed by the $2$-cluster correlation for long-time scale
by machine learning, although the CCE-2 method fails in this case.
This is because there may exist a nonlinear mapping between $\delta S^{(n)}$
and $\tilde{S}_{C_{2}}^{(n)}$ although the linear mapping fails based
the on CCE-2 method in Eq. (\ref{eq:CEM}). The learning procedure
is shown in Fig. \ref{fig:protocol}. We first simulate the quenching
quantum dynamics under the Hamiltonian $\hat{H}$ in Eq. (\ref{eq:Hamiltonian_MBL})
for a certain random configuration $h_{i}$ when the initial state
is prepared as a random product state $|\Psi(0)\rangle=\bigotimes_{i=1}^{N}|m_{i}\rangle$.
Then we calculate the QMI $\delta S^{(n)}$ and irreducible entropy
$\tilde{S}_{C_{L}}^{(n)}$ based on the local correlations $C_{\boldsymbol{\alpha}_{\boldsymbol{i}_{L}}}^{\boldsymbol{i}_{L}}$($L\le2$)
for different quenching times $t$ under various random configurations
and random initial states, which is termed the training set. Finally,
we use the generated irreducible Rényi entropy $\tilde{S}_{C_{L}}^{(n)}$
with $L\le2$ as the input of a multi-layer perceptron neural network
to learn the nonlinear mapping to the QMI(see Fig. \ref{fig:protocol}).
Once the nonlinear mapping is trained, we use it to predict the QMI
by the local correlations generated for random configurations and
initial states totally different from the training set.

To train the QMI via local correlations by the method of MLP network,
we implement a systematic deep learning framework. The activation
function is chosen to be the Gaussian Error Linear Unit (GELU). This
architecture was designed to capture the complex nonlinear mapping
between the local correlations $\tilde{S}_{C_{L}}^{(n)}$($L\le2$)
and the QMI. To capture the broad distribution of QMI, we design a
weighted hybrid loss function which balance overall accuracy and robustness
in capturing weak signals{[}see Supplementary Materials{]}.

\begin{figure}[ptb]
\includegraphics[width=1\columnwidth]{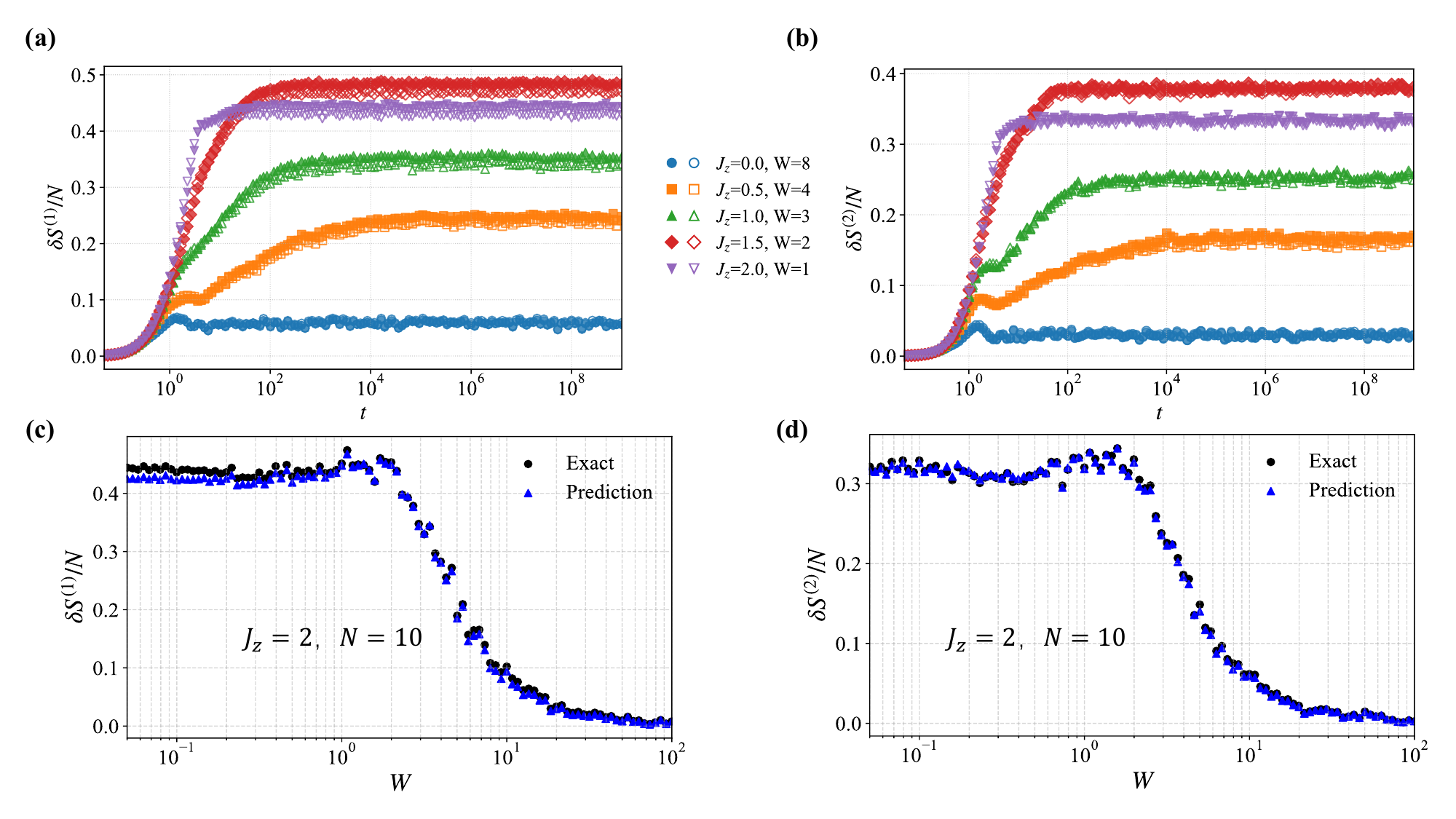} \caption{\textbf{Universal learning of the long-time dynamics of QMI: }The
prediction of long-time dynamics of QMI (a). $\delta S^{(1)}$; (b).
$\delta S^{(2)}$ under different disorder strength $W$ and anisotropy
$J_{z}$ using the universal model trained under the parameters $W=1,J_{z}=2$
and quenching time $t\in[0,200]$. In all these figures, the solid
and empty scatters denotes the the result of exact simulation and
prediction by MLP respectively. Different colors and shape denotes
different parameters as shown in the legend of the graphs. (c) and
(d) shows the prediction of long-time dynamics of QMI $\delta S^{(1)}$
and $\delta S^{(2)}$ respectively for fixed time $t=10^{9}$ and
anisotropy $J_{z}=2$ as a function of disorder strength $W$ using
the universal model trained at the parameters $W=1,J_{z}=2$. The
circle scatters and rectangle denotes the results from exact simulation
and MLP respectively. The size of the system is taken to be $N=10$
and initial state to be Neel states.}
\label{fig:universal} 
\end{figure}

\begin{figure}[ptb]
\includegraphics[width=1\columnwidth]{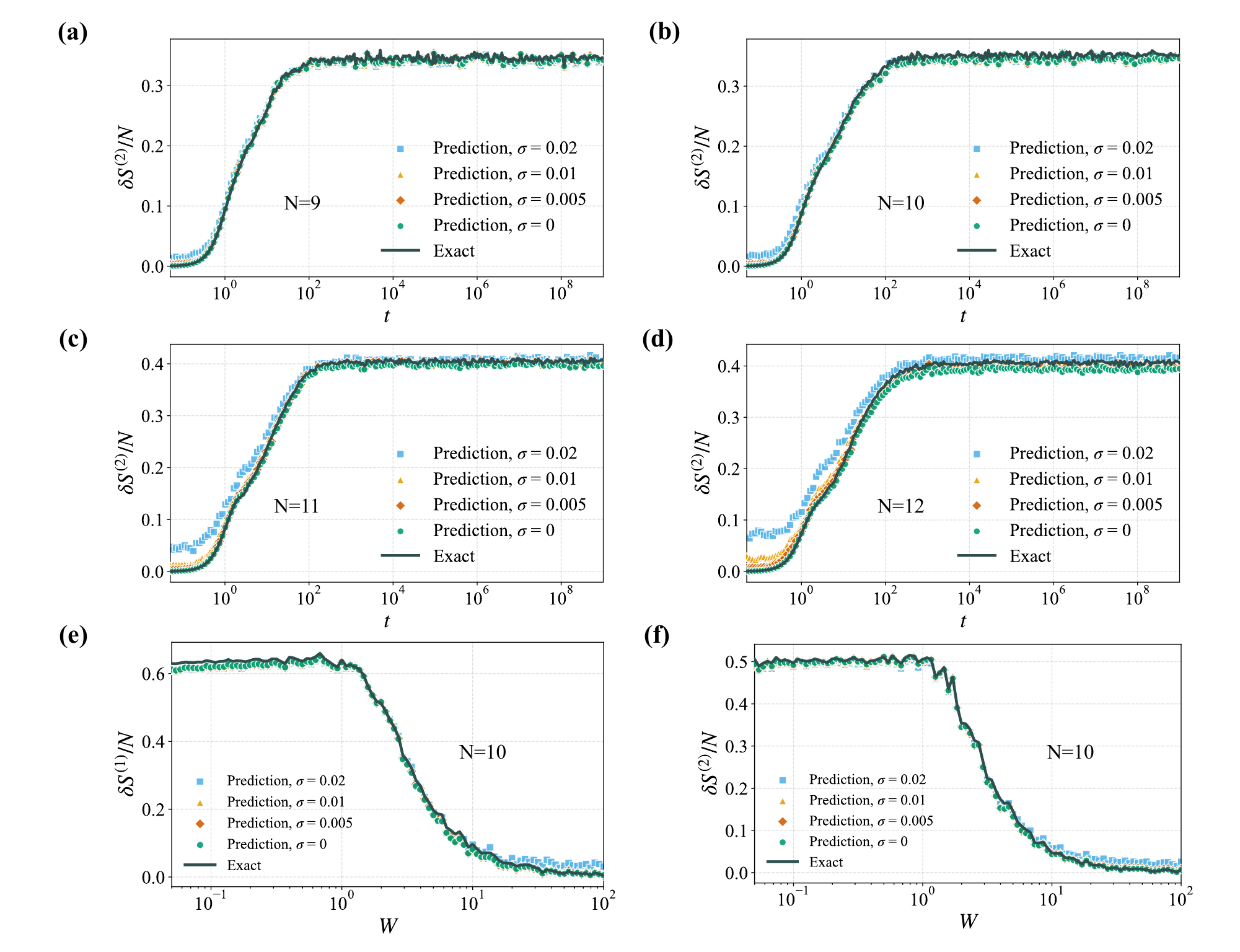} \caption{\textbf{Robustness of the MLP method: }(a) $N=9$, (b) $N=10$, (c)
$N=11$ and $N=12$: Machine learning of of long time dynamic of the
QMI for different strength of the measurement noise and different
system size; (e) and (f): Prediction of QMI $\delta S^{(1)}$ and
$\delta S^{(2)}$for fixed quenching time $t=10^{9}$ as a function
of disorder strength $W$ under different noise strength $\sigma$.
The size of the system is $N=10$. All these prediction use the model
trained at $W=1,J_{z}=1$. }
\label{fig:robust} 
\end{figure}

\textit{Result.---}First, we show that the MLP method precisely reconstructs
the quenching dynamics of the QMI across a broad range of disorder
strength. Since the maximum possible value of QMI is $N$, we predict
the dynamics of $\delta S^{(n)}/N$ for disorder strengths $W=1,3,8$
as shown in Fig.~\ref{fig:Comparision}. Specifically, we take the
initial state to be Neel states $|\Psi\rangle=|01010\cdot\cdot\cdot01\rangle$
without loss of generality{[}see supplementary materials for other
initial states{]}. The upper panels{[}graph (a),(b), (c){]} of Fig.~\ref{fig:Comparision}
show results for a single disorder realization. For comparison, we
also show the corresponding results calculated from CCE method. MLP
method demonstrates obvious advantages over the CCE method. As shown
by the curves in the upper panel of Fig.~\ref{fig:Comparision},
CCE(dotted and dashed line) accurately captures $\delta S^{(n)}/N$
only at short times $t\ll1$ or under strong disorder, whereas our
method maintains accuracy across extended time scales ($\sim10^{9}$)
and all disorder strengths---particularly in the weak-disorder regime.
This limitation of CCE originates from its perturbative nature, which
restricts its validity to weakly correlated systems. As time increases
beyond $1$, correlations propagate to higher orders, developing a
nonperturbative character that renders CCE inadequate even at fourth
order (CCE-4; see supplementary materials). Besides, it should also
be emphasized that our model can predict $\delta S^{(n)}$ for different
$n$(the case $n=1,2$ is demonstrated in Fig. \ref{fig:Comparision}
) while the randomized measurements method in Ref. \cite{BrydgesScience2019}
can only predict the $\delta S^{(2)}$.

The observed tiny discrepancy between the MLP predictions and the
exact simulation results can be attributed to two main factors. First,
it may originate from the inherent limitations of the MLP approach
based on correlations up to the second order. Improving the prediction
accuracy would thus require the inclusion of even higher-order correlations.
Second, the discrepancy could arise from the finite size of the training
dataset. If this is the case, the accuracy should improve with more
training samples. As shown in the Supplementary Material, we indeed
find that the precision continues to increase with larger sample sizes.
This suggests that the finite training data size is likely the dominant
source of the discrepancy. Higher-order correlations could contribute,
but their effect is likely most pronounced in capturing the fine features
of the QMI dynamics.

%predicts the dynamics of exact QMI(black scatters) for small (thermal phase) for long time scale, up to ,  saturates at higher values and the MLP still yields accurate predictions when  approaches (, upper panel) using only second-order correlations. However, the CCE method fails to capture the dynamics of QMI for long time scale[see graph(c) of Fig.] although it shows well performance for short time scale.

For studying the MBL-thermal transition, disorder-averaged QMI is
preferable as it reduces sample-to-sample fluctuations of random configurations~\cite{SinghNJP2016}.
We predict the QMI for individual disorder realizations and then average
them over different configurations to obtain the disorder averaged
QMI. Results in Fig.~\ref{fig:Comparision}{[}lower panel, graphs
(c),(d) and (e){]} demonstrate that the MLP can still provide precise
predictions of the disorder averaged QMI and offers more accurate
predictions than that of single realization of disorder, since the
configuration fluctuation is averaged. For $W=8$ in Fig.~\ref{fig:Comparision}(e)
(MBL phase), the QMI exhibits slow initial growth followed by rapid
oscillations, reflecting correlation propagation inhibited by strong
disorder. The MLP results (blue triangle scatters) faithfully reproduce
both the initial growth dynamics and the oscillatory features of exact
calculations (blue circle scatters), as shown in Fig. \ref{fig:Comparision}
(c). For $W=1$ in Fig.~\ref{fig:Comparision} (f)(thermal phase)
, as shown in Fig. \ref{fig:Comparision} (a), $\delta S^{(n)}/N$
displays rapid initial growth followed by saturation at large values---behavior
that our method can still capture accurately. At intermediate disorder
($W=3$), our approach also provides reliable predictions of the QMI
dynamics, as evidenced in Fig.~\ref{fig:Comparision}(d). These results
clearly illustrate the non-perturbation nature of MLP methods.

Then we demonstrate the universality of our method for the prediction
of the QMI $\delta S^{(n)}$. A universal model means that it can
predict the dynamics of $\delta S^{(n)}$ generated under different
parameters. It indicates that the different quantum states generated
from different quenching process belong to the same complexity class.
If this is true, once the model is trained, we can use a single model
to predict the dynamics for other control parameters. This advantage
will dramatically simplify the prediction process and reduce the time
cost in realistic quantum quenching experiments.

To verify this, we investigate how to predict the long time(up to
$t=10^{9}$) dynamics of QMI $\delta S^{(n)}$ for a broad range of
disorder strengths $W$ based on the model trained from single parameters($W=1$
and $J_{z}=2$). As shown in Fig. \ref{fig:universal}(a), (b), we
plot the exact and predicted $\delta S^{(n)}$ for the case of $n=1,2$
under the long time quenching process. We can see that the MLP model
trained at $W=1,J_{z}=2$ can perfectly predict the long time dynamics
of both $\delta S^{(1)}$ and $\delta S^{(2)}$ for $W=1,2,3,4,8$
and $J_{z}\le2$ from a single model trained in $W=1,J_{z}=2$. The
reason that we choose to model at $W=1,J_{z}=2$ is that the model
should possess enough complexity to encompass other cases. In summary,
these results, in one aspect, show that the local second order correlations
are enough to predict the complex quantum state even in the thermal
phase ($W=1,2,3<W_{\mathrm{c}}$), where the QMI has scrambled to
$\sim0.6N$, a highly entanglement state which is usually expected
to fail to describe by only second order correlations. In the other
aspect, it implies an underlying similarity among these nonequilibrium
quantum states, despite their preparation at different disorder strengths.

To show the power of the method, we also predict QMI at a fixed long-time
scale $t=10^{9}$ as a function of disorder strength $W$, whose performance
has been previously used to identify critical phenomena near the many-body
localization (MBL) transition~\cite{SinghNJP2016}. Remarkably, this
prediction is made despite the training data being restricted to the
much shorter time interval $t\in[0,200]$. As shown in Fig.~\ref{fig:universal}(c,d),
our model accurately captures the long-time behavior of $\delta S^{(n)}$
versus disorder strength, even though the local second-order correlations
have completely evolved during this extended duration, which means
the trained nonlinear function indeed captures the universal features
of the complex state in one-dimensional XXZ model. This predictive
power enables us to extract the critical point of the MBL-Thermal
transition through the reconstruction of the QMI in quench dynamics
and is particularly significant because conventional approaches require
eigenstate properties that are experimentally inaccessible\cite{KhemaniPRX2017}.

Finally, we examine the robustness of our prediction model against
experimental measurement noise. In realistic settings, the noise in
correlation measurements---determined by the number of measurement
shots $M$ and the signal-to-noise ratio (SNR)---introduces errors
in the predicted QMI. To address this, we add Gaussian noise with
standard deviation $\sigma$ to the expectation values and second-order
correlations, and use the noisy data to predict the QMI. Since the
noisy data would change the positivity of density matrix and hence
be invalid to compute the inputs $\tilde{S}_{C_{1}}^{(n)}$ and $\tilde{S}_{C_{2}}^{(n)}$
of the MLP, we correct the noisy data via the optimization method
shown in the Supplementary Materials. Using the corrected data, the
prediction results are shown in Fig.~\ref{fig:robust}. We observe
that increasing the system size from $N=9$ to $N=12$ does not significantly
degrade the prediction accuracy, suggesting that the trained model
is robust to measurement noise, which remains accurate up to $\sigma\approx0.01$.
Based on the noise strength, we provide an estimation the possibility
of experimental realization in different quantum platforms. By the
central limit theorem, this noise level $\sigma\approx0.01$ corresponds
to a measurement budget of $M\approx1/\sigma^{2}\approx10^{4}$ per
second-order correlation, assuming perfect readout fidelity. The total
number of measurements is then estimated as $9N(N-1)M/2\approx6\times10^{6}$($N\sim12$).
In superconducting qubits, this corresponds to a total measurement
time of approximately $\sim3~\mathrm{s}$ for each time point and
single configuration using the characteristic quenching time $10^{2}/J\sim4\mathrm{\mu s}$\cite{PitaVidalNP2024}($J$
be the coupling strength of superconducting qubits) and a fast single-shot
readout time of $100~\mathrm{ns}$ in superconducting qubits system\cite{WalterPRApplied2017,SunadaPRApplied2022,GunyhoNE2024}.

%In nuclear spin bath of nitrogen vacancy center, the total time budget is estimated to be for enriched  nuclear spin bath(abundance ) with coupling to be .

\textit{Conclusion and Discussion.---}In this Letter, we introduce
a machine learning framework based on MLP to efficiently predict the
QMI in quench dynamics of the one-dimensional disordered XXZ spin
chain, using only experimentally accessible second-order correlations
and avoiding full quantum state tomography. Our approach successfully
captures the non-equilibrium dynamics of the QMI across both many-body
localized and thermalizing regimes.

Notably, the MLP-based method significantly outperforms conventional
CCE techniques, especially at long times and in regimes of strong
entanglement, where perturbative expansions break down. We further
demonstrate the universality of the learned model: a network trained
at a specific disorder strength accurately predicts QMI dynamics over
a wide range of disorders and times far beyond those seen during training.
This suggests that the mapping from low-order correlations to non-local
entropies exhibits a universal character across different dynamical
phases. Moreover, the method proves robust to realistic measurement
noise, indicating practical feasibility in experimental platforms
such as superconducting qubits and ion traps with finite sampling
resources.

Our results open a promising route toward the efficient and scalable
characterization of non-local quantum information in many-body systems.
More broadly, this data-driven approach illustrates how machine learning
can uncover hidden universal relations between local observables and
complex quantum properties in non-equilibrium dynamics, providing
a powerful tool for studying quantum dynamics, phase transitions,
and entanglement structures in regimes where traditional methods are
intractable.

Future directions include extending the approach to higher dimensions
and long-range interaction models, open quantum systems, other non-equilibrium
dynamics, Fisher information, and out of time correlations\cite{MacFarlanePTORSL2003},
as well as combining it with advanced experimental techniques such
as tensor networks\cite{LanyonNP2017}, randomized measurements\cite{BrydgesScience2019},
or shadow tomography\cite{HuangNP2020} to further enhance data efficiency.

\textit{Acknowledge.---} P. W. is supported by the National Natural
Science Foundation of China(Grant No. 12475012, Grant No. 62461160263),
the Guangdong Provincial Quantum Science Strategic Initiative (Grant
GDZX2403009, No. GDZX2303005) and the Quantum Science and Technology-National
Science and Technology Major Project of China (Project 2023ZD0300600).
H. L. is supported by the National Natural Science Foundation of China
under Grant No. 62276171, the Guangdong Basic and Applied Basic Research
Foundation, China, under Grant No. 2024A1515011938, and the Shenzhen
Fundamental Research-General Project, China, under Grant No. JCYJ20240813141503005.

% === 补充材料声明放在这里 ===
\par\noindent\textbf{Supplementary Material}
\par\noindent Detailed calculations, additional data figures, and methods are provided in the Supplementary Material \cite{supp}.

%\bibliography{Deep_Learning}

%apsrev4-2.bst 2019-01-14 (MD) hand-edited version of apsrev4-1.bst
%Control: key (0)
%Control: author (8) initials jnrlst
%Control: editor formatted (1) identically to author
%Control: production of article title (0) allowed
%Control: page (0) single
%Control: year (1) truncated
%Control: production of eprint (0) enabled
\providecommand{\noopsort}[1]{}\providecommand{\singleletter}[1]{#1}%

\end{document}

\title{Supplementary information for ``Universal learning of nonlocal entropy via local correlations in non-equilibrium quantum states''}
\today

\author{Hao Liao}
\affiliation{National Engineering Laboratory on Big Data System Computing Technology, Guangdong Province Engineering Center of
China-made High Performance Data Computing System, College of Computer Science and Software Engineering, Shenzhen
University, Shenzhen 518060, China}

\author{Xuanqin Huang}
\affiliation{National Engineering Laboratory on Big Data System Computing Technology, Guangdong Province Engineering Center of
China-made High Performance Data Computing System, College of Computer Science and Software Engineering, Shenzhen
University, Shenzhen 518060, China}

\author{Ping Wang}
\email{wpking@bnu.edu.cn}
\affiliation{Faculty of Arts and Sciences, Beijing Normal University, Zhuhai 519087, China}

\maketitle

\tableofcontents

\subsection{The CCE method}

The QMI can be written as 
\[
\delta S^{(n)}=S_{A}^{(n)}+S_{B}^{(n)}-S^{(n)}.
\]
Then we expand the entropy by cluster expansion method as following
\begin{alignat*}{1}
 & S_{A}^{(n)}=\sum_{M=1}^{N_{A}}\sum_{C_{M}\subseteq A}\tilde{S}_{C_{M}}\\
 & S_{B}^{(n)}=\sum_{M=1}^{N_{B}}\sum_{C_{M}\subseteq B}\tilde{S}_{C_{M}}\\
 & S_{}^{(n)}=\sum_{M=1}^{N}\sum_{C_{M}\subseteq A+B}\tilde{S}_{C_{M}}.
\end{alignat*}
As a result, we have 
\[
\delta S^{(n)}=-\sum_{M=1}^{N}\sum_{C_{M}\cap A,C_{M}\cap B\ne\phi}\tilde{S}_{C_{M}}
\]

\subsection{Estimation of correlations via noisy data}

The density matrix of the two spin system can be written as

\begin{alignat*}{1}
 & \hat{\rho}=\frac{1}{4}\left(1+\sum_{\alpha}C_{\alpha}^{(1)}\hat{\sigma}_{\alpha}+\sum_{\alpha}C_{\alpha}^{(2)}\hat{\sigma}_{\alpha}+\sum_{i}C_{\alpha\beta}^{(1,2)}\hat{\sigma}_{\alpha}\hat{\sigma}_{\beta}\right)
\end{alignat*}
$C_{\alpha}^{(1)},C_{\beta}^{(2)},C_{\alpha\beta}^{(1,2)}$ are the
realistic correlations and the experimentally measured correlation
is denoted as $T_{\alpha}^{(1)}$, $T_{\beta}^{(2)}$,$T_{\alpha\beta}^{(1,2)}$.
To estimate the realistic correlations $T_{\alpha}^{(1)},T_{\beta}^{(2)},T_{\alpha\beta}^{(1,2)}$,
we should minimize the loss function 
\begin{equation}
\Delta=\sum_{\alpha,i}\frac{\left(T_{\alpha}^{(i)}-C_{\alpha}^{(i)}\right)^{2}}{\sigma_{1}^{2}}+\sum_{\alpha\beta}\frac{\left(T_{\alpha\beta}^{(1,2)}-C_{\alpha\beta}^{(1,2)}\right)^{2}}{\sigma_{2}^{2}},
\end{equation}
where $\sigma_{1},\sigma_{2}$ is the measurement noise of first order
correlation $T_{\alpha}^{(1)},T_{\alpha}^{(2)}$ and second order
correlation $T_{\alpha\beta}^{(1,2)}$, respectively, while simultaneously
maintaining the positivity of the density matrix. To achieve this
goal, we diagonalize the density matrix as following

\begin{equation}
\hat{\rho}=U\begin{pmatrix}\lambda_{1} & 0 & 0 & 0\\
0 & \lambda_{2} & 0 & 0\\
0 & 0 & \lambda_{3} & 0\\
0 & 0 & 0 & \lambda_{4}
\end{pmatrix}U^{\dagger}
\end{equation}
which are quantified by 36 parameters $\lambda_{i}$, $\mathrm{Re}U_{ij}$
, and $\mathrm{Im}U_{ij}$ ($U_{ij}$ has the number $2N^{2}=32$).
To maintain the positivity and normalization of the probability of
$\hat{\rho}$, we apply the constraints to the eigen values 
\begin{alignat}{1}
 & \lambda_{1}+\lambda_{2}+\lambda_{3}+\lambda_{4}=1,\lambda_{i}\ge0\label{eq:constraint_positive}
\end{alignat}
Besides, due to the unitary properties $UU^{\dagger}=1$, another
$N^{2}=16$ constraints 
\begin{align}
 & \left(\mathrm{Re}U_{ik}\right)^{2}+\left(\mathrm{Im}U_{ik}\right)^{2}=1,\label{eq:constraint_unitary}\\
 & \mathrm{Re}U_{ik}\mathrm{Re}U_{jk}+\mathrm{Im}U_{ik}\mathrm{Im}U_{jk}=0,\mathrm{Im}U_{ik}\mathrm{Re}U_{jk}-\mathrm{Re}U_{ik}\mathrm{Im}U_{jk}=0,\nonumber 
\end{align}
where the first equations give $N$ constraints while the second give
$N(N-1)$ constraints.

Another constraints are the phase redundancy of eigen vectors of $\hat{\rho}$
due to the density matrix $\hat{\rho}$ is unchanged if a global phase
is applied to the four eigen vectors of $\hat{\rho}$. This redundancy
can be removed by four constraints 
\begin{equation}
\mathrm{Re}U_{1k}>0,\mathrm{Im}U_{1k}=0\label{eq:constraint_phase_redandancy}
\end{equation}
As a result, the independents number of the parameters are only $36-1-16-4=15$.
To keep the numerical stability, the constraints Eq. (\ref{eq:constraint_phase_redandancy})
is usually implemented by the following methods 
\[
\mathrm{Re}w_{i}U_{ij}>0,\mathrm{Im}\sum_{i}w_{i}U_{ij}=0
\]
where $w_{i}$ is smooth weights depends on $\left|U_{ij}\right|^{2}$to
avoid the extreme case that $\mathrm{Re}U_{1k}=0$.

\subsection{Optimization of the hyperparameters of the MLP}

This section outlines the systematic optimization of the hyperparameters
for the multilayer perceptron (MLP). Hyperparameters are configuration
variables that control the model’s architecture---such as the number
of layers and neurons---as well as the training strategy, including
the learning rate and batch size, and regularization methods like
dropout rate and weight decay. In contrast to model parameters (weights
and biases), which are learned during training, hyperparameters must
be set a priori and critically influence the model’s learning capacity,
convergence behavior, and generalization ability. Finding their optimal
values is nontrivial and often necessitates automated search techniques.
We therefore adopt a Bayesian optimization approach, implemented via
the Optuna library, to efficiently locate the optimal hyperparameter
configuration within the search space defined in Table .\ref{tab:Hyperparameter}

The optimization focused on structural parameters---specifically,
the number of hidden layers and neurons per layer---as well as training
settings such as the learning rate, batch size, and number of epochs,
along with regularization mechanisms including the dropout rate and
weight decay. To efficiently sample this high-dimensional space, we
employed the Tree-structured Parzen Estimator (TPE) sampler integrated
with the Hyperband pruning strategy. A total of 150 optimization trials
were performed to balance the trade-off between exploration and computational
expense.

\begin{table*}[htbp]
\centering \caption{Hyperparameter search space for Bayesian optimization}
\label{tab:Hyperparameter} \setlength{\tabcolsep}{5pt} %列间距
\global\long\def\arraystretch{1.4}%
%
%
% 行距
\begin{tabular}{@{}ll@{}}
\hline 
Parameter  & Range/Options \tabularnewline
\hline 
Batch size  & $\{64,128,256,\mathbf{512}\}$ \tabularnewline
Activation function  & $\{\mathrm{ReLU},\mathrm{LeakyReLU},\mathrm{Tanh},\mathrm{GELU}\}$ \tabularnewline
Hidden layers  & $[2,5]$ \tabularnewline
Neurons per hidden layer  & $\{16,\mathbf{32},64,128,256,512\}$ \tabularnewline
Weight decay (AdamW)  & $10^{-7}$ to $10^{-3}$ \tabularnewline
Learning rate  & $10^{-6}$ to $10^{-2}$ \tabularnewline
\hline 
\end{tabular}
\end{table*}

The optimized hyperparameter is shown as following: 
\begin{enumerate}
\item {Hidden layers: 512 and 32 neurons for the first and second hidden
layers respectively} 
\item {Activation function: GELU(Gaussian Error Linear Unit) is defined
as $\ensuremath{\text{GELU}(x)=x\left[1+\text{erf}\left(x/\sqrt{2}\right)\right]/2}$,
where $\text{erf}$ is error function} 
\item {Dropout: disabled ($p=0$)} 
\item {Learning rate: $10^{-3}$} 
\item {Batch size: 256} 
\item {Weight decay: $4\times10^{-3}$} 
\end{enumerate}

\subsection{Hybrid MSE-MAE loss function}

Then the training set is divided into many batches to train the MLP
and the batch size is denoted as $N_{b}$. We use $\mathbf{y}_{\text{true}}=\{y_{\text{true},1},y_{\text{true},2},\cdot\cdot\cdot,y_{\text{true},N_{b}}\}$
to denote the list of the exact simulated $\delta S^{(n)}$ of the
training set in each batch. Since $\delta S^{(n)}$ is distributed
from $0$ to its maximum, we design a weighted hybrid loss function:
\begin{equation}
\begin{aligned} & \mathcal{L}(\mathbf{y}_{\text{pred}},\mathbf{y}_{\text{true}})=\frac{(1-\eta)\mathcal{D}^{\mathrm{S}}(\mathbf{y}_{\text{pre}}^{\mathrm{S}},\mathbf{y}_{\text{true}}^{\mathrm{S}})+\eta\mathcal{D}^{\mathrm{L}}(\mathbf{y}_{\text{pre}}^{\mathrm{L}},\mathbf{y}_{\text{true}}^{\mathrm{L}})}{N_{b}}\end{aligned}
\label{eq:custom_loss}
\end{equation}
where the data is divided into two groups: $\mathbf{y}_{\text{true}}^{\mathrm{S}}$for
values below a threshold $0.03$, and $\mathbf{y}_{\text{true}}^{\mathrm{L}}$
for those above it. $\mathbf{y}_{\text{pre}}^{\mathrm{L}/\mathrm{S}}$
denotes the corresponding predicted value of them. In this loss function,
we use the squared error $\mathcal{D}^{\mathrm{L}}(\mathbf{y}_{\text{pre}}^{\mathrm{L}},\mathbf{y}_{\text{true}}^{\mathrm{L}})\equiv\sum_{i=1}(y_{\text{pre},i}^{\mathrm{L}}-y_{\text{true},i}^{\mathrm{L}})^{2}$
for the larger values, while the absolute error $\mathcal{D}^{\mathrm{S}}(\mathbf{y}_{\text{pre}}^{\mathrm{S}},\mathbf{y}_{\text{true}}^{\mathrm{S}})\equiv\sum_{i=1}|y_{\text{pre},i}^{\mathrm{S}}-y_{\text{true},i}^{\mathrm{S}}|$
is adopted for smaller values. This approach emphasizes the squared
error when targets exceed the threshold and prioritizes the absolute
error for weaker signals, thus balancing overall accuracy and robustness
in capturing weak signals.

\subsection{Training of the multi-layer perception model}

This note supplements the implementation details of the MLP model
presented in the main text. The model is constructed using the PyTorch
framework. Input features comprise the irreducible Rényi entropies,
$\tilde{S}_{C_{1}}^{(n)}$ and $\tilde{S}_{C_{2}}^{(n)}$, for all
single-site and two-site clusters. These inputs are standardized to
the interval $(0,1)$ via conventional normalization to ensure uniform
numerical scaling. The complete dataset is randomly partitioned into
$80\%$ for training and $20\%$ for testing. During training, we
employ a batch size of $256$ and set the maximum number of epochs
to $100$. An early stopping criterion (patience of $10$ epochs,
minimum improvement of $10^{-5}$) is implemented to mitigate overfitting
and enhance generalization. The training process is optimized using
an MSE--MAE hybrid loss function. Based on the empirical distribution
of the data, a threshold of $0.03$ is chosen to distinguish between
small and large QMI values, with a weighting factor $\eta=0.7$ applied
in the hybrid loss. The final network architecture is derived through
an iterative refinement process starting from a minimal baseline model.

\subsection{Model evaluation}

The model performance is evaluated using the coefficient of determination
($R^{2}$), given by 
\begin{equation}
R^{2}=1-\frac{\sum_{i=1}^{N}(y_{\text{true},i}-y_{\text{pred},i})^{2}}{\sum_{i=1}^{N}(y_{\text{true},i}-\bar{y}_{\text{true}})^{2}}\label{eq:r_squared}
\end{equation}
where $y_{\text{pred},i}$ is the predicted value, $y_{\text{true},i}$
denotes the true value, and $N$ represents the total number of data
points. The closer of $R^{2}$ to $1$ indicates the better of the
performance of the model.

\subsection{The effect of training scale}

\begin{figure}[ptb]
\includegraphics[width=1\columnwidth]{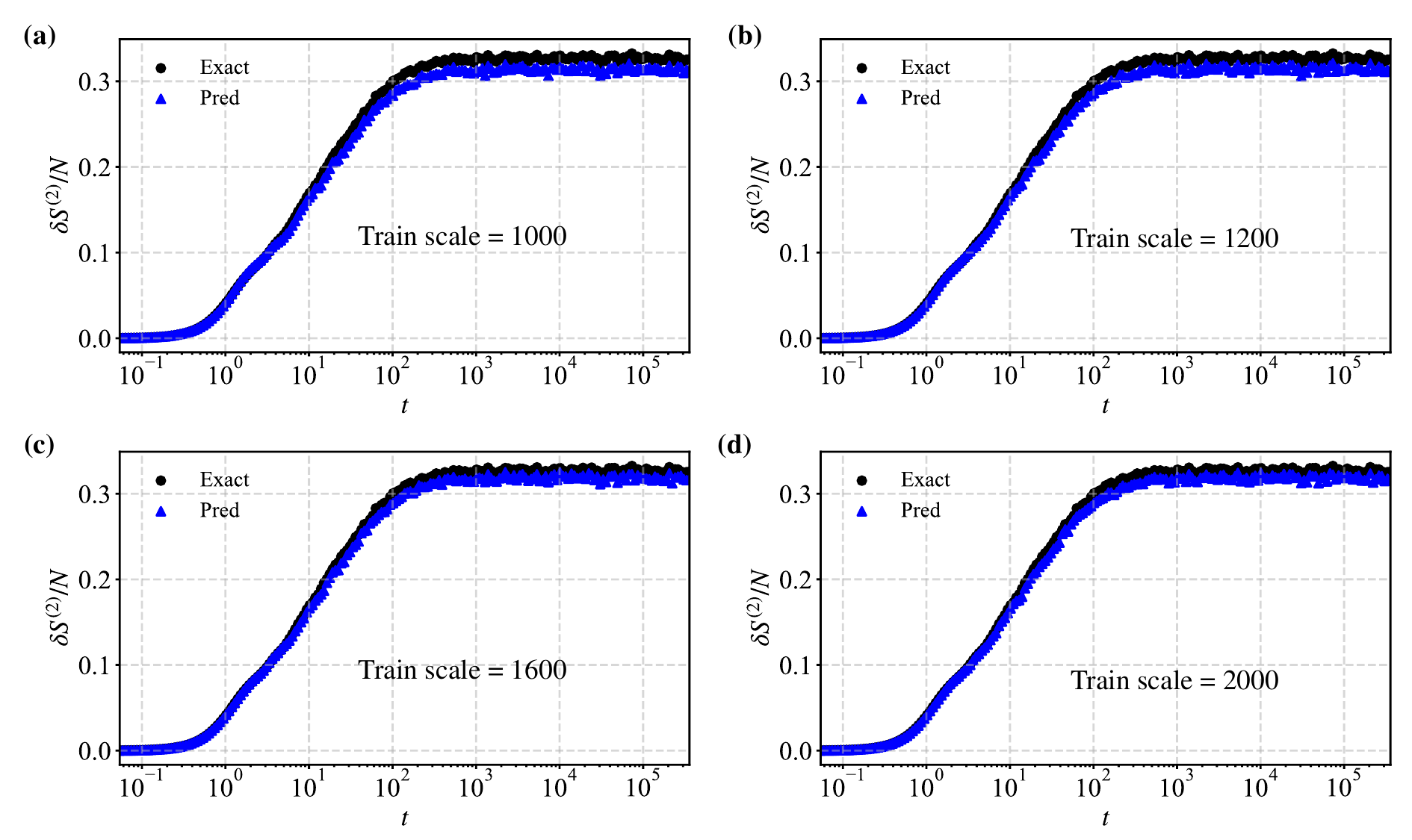} \caption{\textbf{Impact of training set scale on the model's prediction performance.
}The figure shows the averaged prediction results over 100 random
initial states with parameters $N=12,J_{z}=1,n=2,W=2$. The horizontal
axis is time, and the vertical axis is the normalized second-order
irreducible entropy. The four subplots correspond to different training
set scales: (a) $1000$, (b) $1200$, (c) $1600$, and (d) $2000$
samples. The circles represent the exact values, while the triangles
represent the predicted values. It can be seen that, within this parameter
range, increasing the scale of the training set significantly improves
the agreement between the predicted and exact values, demonstrating
a positive correlation.}
\label{fig:effect_sample} 
\end{figure}

In Fig. \ref{fig:effect_sample}, we have investigate the effect of
the sample scale on the training performance. As the increasing of
training performance, the behavior performance improves significantly.

\begin{figure}[htbp]
\centering \includegraphics[width=0.5\columnwidth]{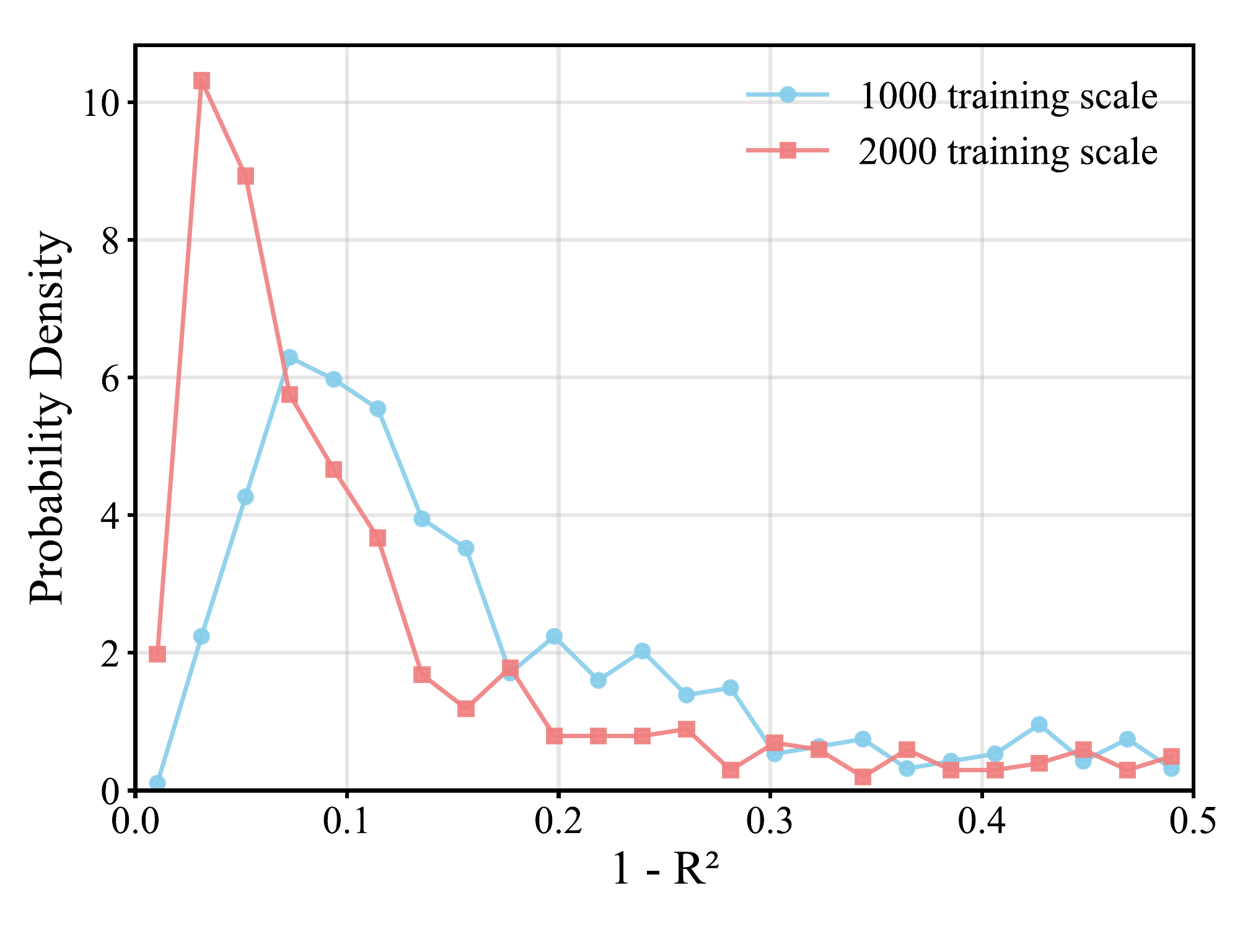} % 调整为80%栏宽
\caption{Probability density of the error $1-R^{2}$ of $\delta S^{(2)}$ for
different training scales. The parameters are set to $J_{z}=1$, $W=3$.}
\label{fig:statistic_error} 
\end{figure}

We show the probability density of the fitting errors, defined as
$1-R^{2}$, for different training scales in Fig.$~$$\ref{fig:statistic_error}$.
As seen from the figure, the peak of the probability density clearly
shifts to lower values with increasing training scale, indicating
a reduction in the fitting error.

\subsection{The effect of noise on CCE-2 method}

\begin{figure}[ptb]
\includegraphics[width=1\columnwidth]{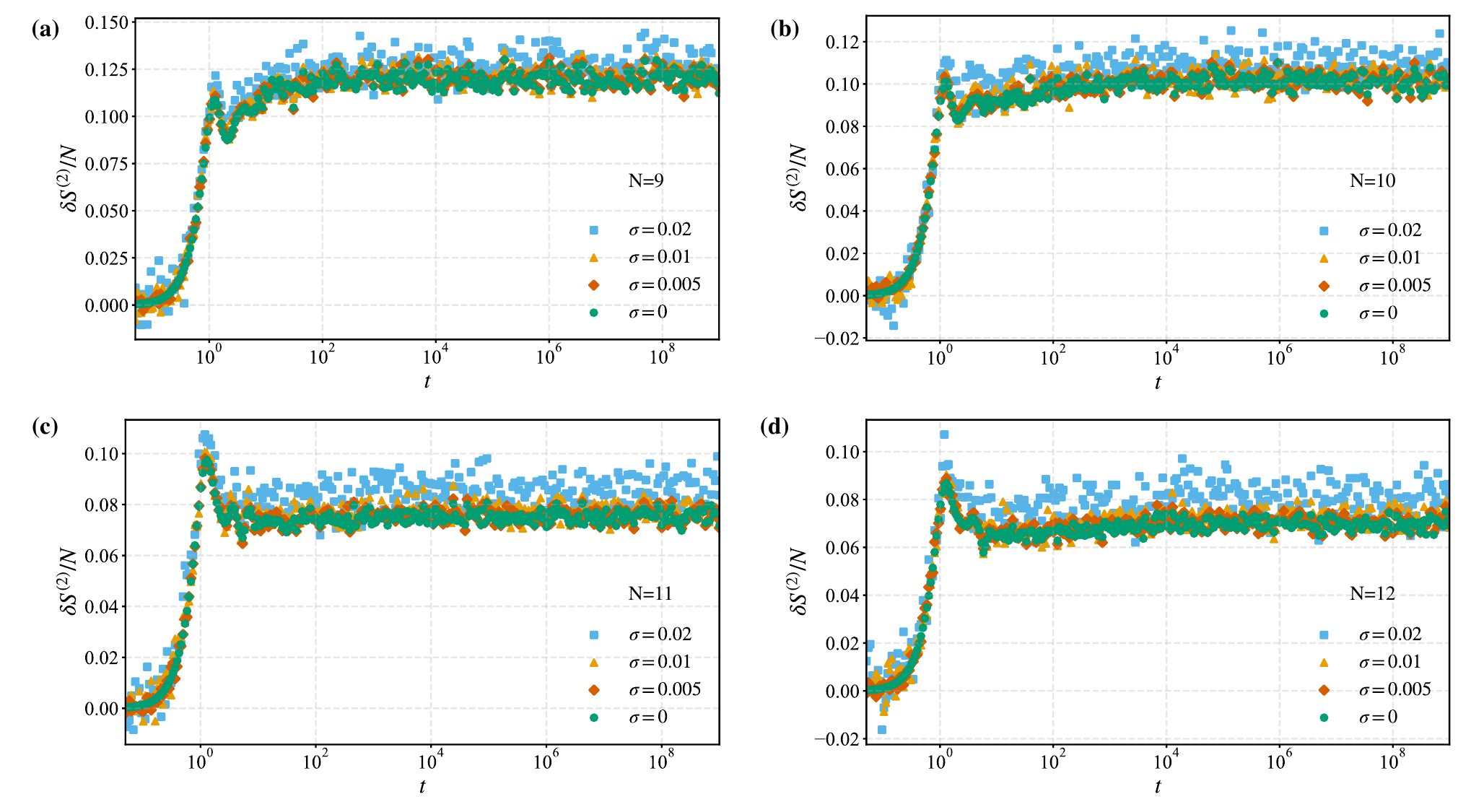} \caption{\textbf{Effect of noise on the second-order truncated entropy (CCE-2):}
Subplots show the dynamics of CEE-2 for different system sizes ($N=9,10,11,12$)
with parameters $J_{z}=1,n=2,W=2$, Gaussian noise ($\sigma=0,0.005,0.01,0.02$)
to the correlation functions. Larger system sizes (e.g., $N=11,12$)
exhibit more pronounced initial oscillations under strong noise.}
\label{fig:noise_cce} 
\end{figure}

In Fig. \ref{fig:noise_cce}, we have investigate the effect of the
measurement noise on prediction precision of MLP.

\subsection{Prediction of QMI for other initial states}

\begin{figure}[ptb]
\includegraphics[width=1\columnwidth]{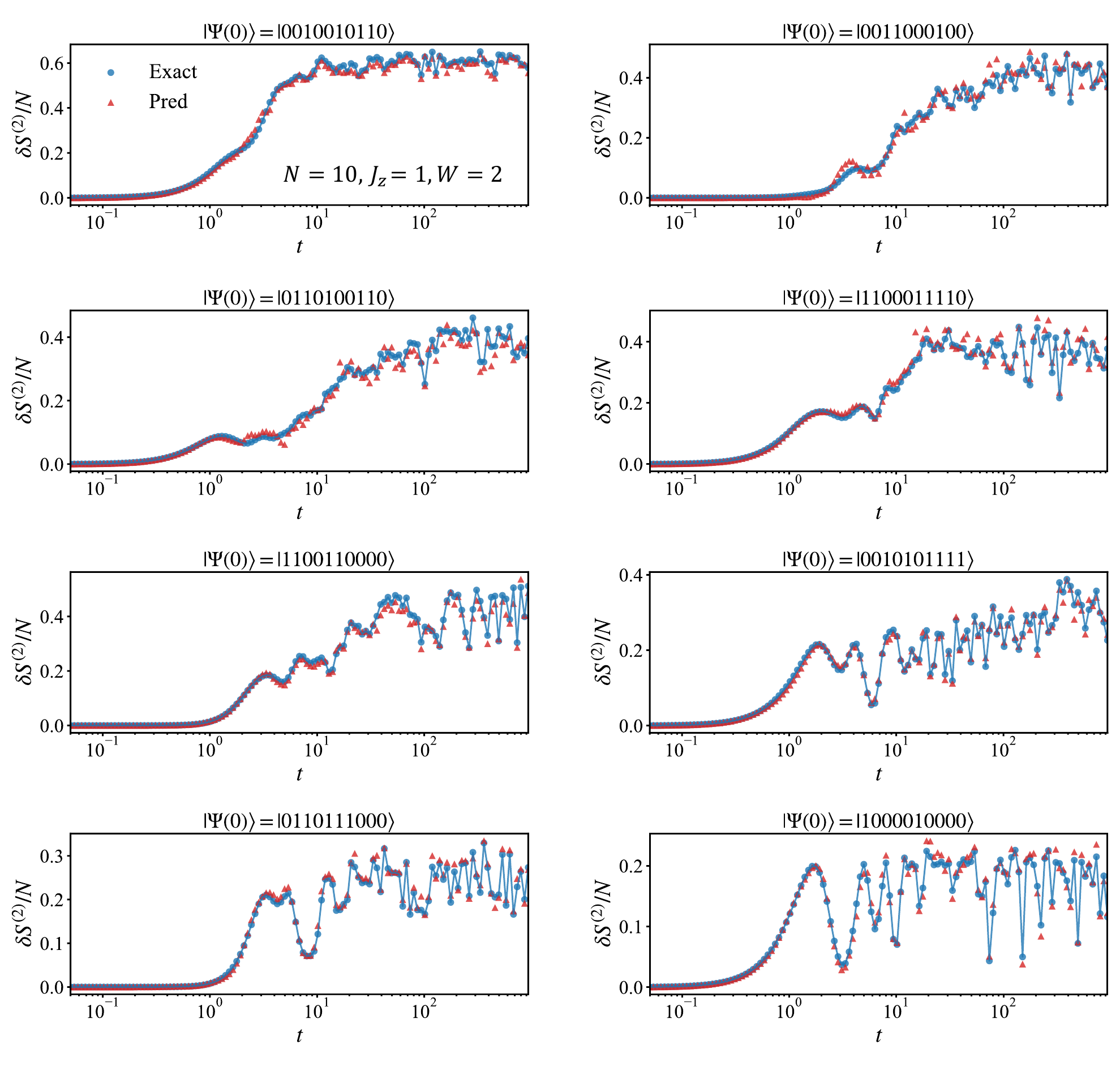} \caption{\textbf{Prediction of QMI for different initial state:} }
\label{fig:prediction_other_states} 
\end{figure}

In Fig. \ref{fig:prediction_other_states}, we show the prediction
of $\delta S^{(2)}$ for various initial states.

\subsection{Prediction of QMI via CCE-4}

\begin{figure}[ptb]
\includegraphics[width=1\columnwidth]{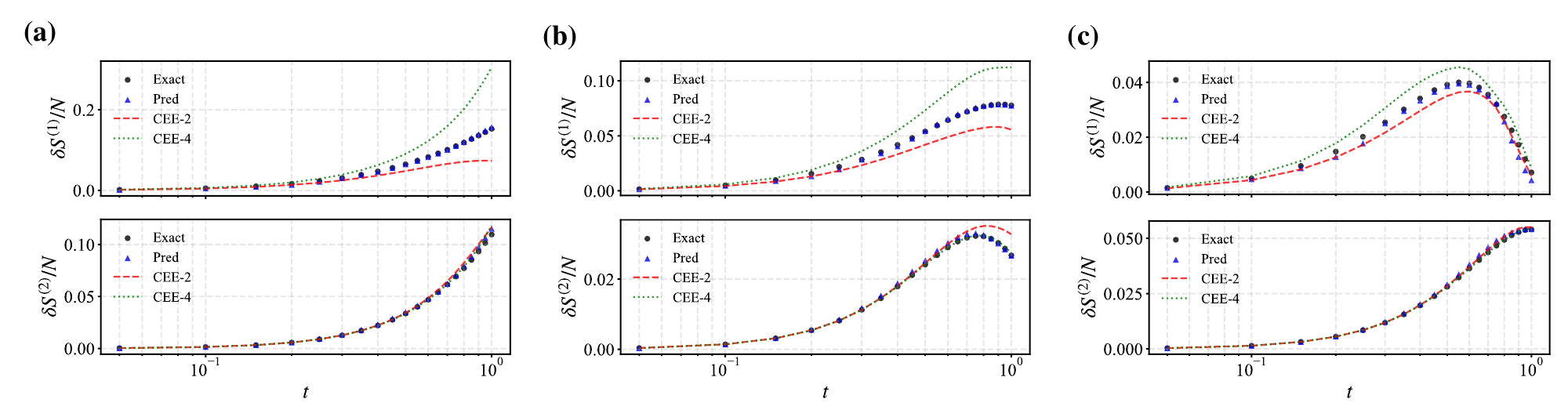} \caption{\textbf{The comparison between machine learning and CCE method:}}
\label{fig:cce-4} 
\end{figure}

In Fig. \ref{fig:cce-4}, we show the comparison between the prediction
of $\delta S^{(n)}$ via machine learning and CCE-4. In this figures,
we can see that the CCE-4 breaks for long time scale especially for
the case of low disorder strength.

% The \nocite command causes all entries in a bibliography to be printed out
% whether or not they are actually referenced in the text. This is appropriate
% for the sample file to show the different styles of references, but authors
% most likely will not want to use it.

%\bibliography{Deep_Learning}
 % Produces the bibliography via BibTeX.

%apsrev4-2.bst 2019-01-14 (MD) hand-edited version of apsrev4-1.bst
%Control: key (0)
%Control: author (8) initials jnrlst
%Control: editor formatted (1) identically to author
%Control: production of article title (0) allowed
%Control: page (0) single
%Control: year (1) truncated
%Control: production of eprint (0) enabled
\providecommand{\noopsort}[1]{}\providecommand{\singleletter}[1]{#1}%